\documentclass[10pt,final,journal,twocolumn]{IEEEtran}
\usepackage{amsmath}
\usepackage{subfigure}
\usepackage{graphicx}
\usepackage{latexsym}
\usepackage{amssymb,amsbsy,verbatim,array}
\usepackage{epstopdf}
\usepackage{color}
\usepackage[T1]{fontenc}
\usepackage{array}
\usepackage{booktabs}

\usepackage[colorlinks]{hyperref}

\newtheorem{remark}{Remark}

\def\rb{\rho_1}
\def\ra{\rho_0}

\def\sd{=1,\cdots, D}

\everymath{\displaystyle}\begin{document}
\title{FAS-RIS: A Block-Correlation Model Analysis}

\author{Xiazhi Lai, Junteng Yao, Kangda Zhi, Tuo Wu,\\
David Morales-Jimenez, \emph{Senior Member}, \emph{IEEE}, and Kai-Kit Wong, \emph{Fellow}, \emph{IEEE}
\thanks{(\textit{Corresponding author: Tuo Wu.})}
\thanks{X. Lai is with the School of Computer Science, Guangdong University of Education, Guangzhou, Guangdong, China (E-mail: xzlai@outlook.com).}
\thanks{J. Yao is with the Faculty of Electrical Engineering and Computer Science, Ningbo University, Ningbo 315211, China (E-mail: yaojunteng@nbu.edu.cn).}
\thanks{K. Zhi and T. Wu are with the School of Electronic Engineering and Computer Science at Queen Mary University of London, London E1 4NS, U.K. (Email: \{tuo.wu, k.zhi\}@qmul.ac.uk).}
\thanks{D. Morales-Jimenez is with Department of Signal Theory, Networking and Communications, Universidad de Granada, Granada 18071, Spain (Email: dmorales@ugr.es).}
\thanks{K.-K. Wong is with the Department of Electronic and Electrical Engineering, University College London, WC1E 6BT London, U.K., and also with the Yonsei Frontier Laboratory and the School of Integrated Technology, Yonsei University, Seoul 03722, South Korea (e-mail: kai-kit.wong@ucl.ac.uk).}
\vspace{-8mm}
}
\markboth{}
{Lai \MakeLowercase{\textit{et al.}}: }

\maketitle

\thispagestyle{empty}

\begin{abstract}
In this correspondence, we analyze the performance of a reconfigurable intelligent surface (RIS)-aided communication system that involves a fluid antenna system (FAS)-enabled receiver. By applying the central limit theorem (CLT), we derive approximate expressions for the system outage probability when the RIS has a large number of elements. Also, we adopt the block-correlation channel model to simplify the outage probability expressions, reducing the computational complexity and shedding light on the impact of the number of ports. Numerical results validate the effectiveness of our analysis, especially in scenarios with a large number of RIS elements.
\end{abstract}


\section{Introduction}
\IEEEPARstart{F}{luid} antenna system (FAS) has emerged as a crucial technology for next-generation wireless communications. This importance arises because traditional fixed antenna techniques, e.g., multiple-input-multiple-output (MIMO), are constrained by the physical  size of wireless devices, whereas FAS enables flexible switching of finely-positioned elements to the most desirable port, e.g., \cite{FAS20,XLai23,JZheng24, HXu24, HXu23, MFAS23}.  In practice, FAS can be implemented via pixel-based \cite{Rodrigo14} or liquid metal structures \cite{Huang21}. The optimal port can be determined using learning-based methods exploiting spatial correlation \cite{Chai22,Waqar23}. FAS has made an active research topic recently, see e.g., \cite{Vega-2023,Vega-2023-2,Alvim-2023,Psomas-dec2023,Dai-2023,zhu2023index,Xu-2024}. An overview article can be found in \cite{Zhu-Wong-2024}.

To study the achievable performance of FAS-enabled systems,  accurate while tractable channel models are necessary. Considering this, a simplified channel model was originally used in \cite{FAS21,FAS22}. This model approximates the effect of port correlation using a single correlation parameter. A more accurate eigenvalue-based channel model was subsequentially proposed in \cite{Khammassi23}. Due to mathematical difficulty, performance analysis based on this model is, however, a challenge. This was recently overcome by the new block-correlation model developed in \cite{BC24}, which is favorable for insightful analysis of FAS-assisted communications.

On the other hand, research on reconfigurable intelligent surface (RIS) is going strong. RIS is particularly important in deep-fading scenarios where obstacles such as thin walls may obstruct the direct link between the transmitter and receiver \cite{TWu1,TWu2,XHu24}. RIS in a sense re-establishes the line-of-sight (LoS) link between the transmitter and receiver when it is blocked by modifying the phases of the reflected signals. Recent efforts have seen the benefits of RIS in terms of coverage probability in \cite{Yang20}, and \cite{Gan21} also considered the use of RIS in the presence of multiple users, in terms of ergodic capacity.

Integrating the capabilities of FAS and RIS presents a robust approach to enhance wireless communication performance, so the synergy of FAS and RIS offers a promising direction for advancing communication systems. However, the performance of FAS-RIS is not well understood, and remains unexplored. Motivated by this, this correspondence explores a RIS-assisted communication system that includes a FAS-enabled receiver. Here, FAS is used to improve the quality of received signals, while RIS is employed to re-establish the LoS link when it is originally blocked. Using the central limit theorem (CLT), we derive the system outage probability when the number of RIS elements is large. Also, the block-correlation approximation technique in \cite{BC24} is adopted to achieve simplified expressions for the outage probability. Our numerical results validate our analysis and reveal great potential of FAS and RIS.

\emph{Notations:} $X\sim\mathcal{CN}(\alpha,\beta)$ represents a complex Gaussian random variable (RV) with mean $\alpha$ and variance $\beta$. $\mathbf{E}(\cdot)$ and $\mathbf{Var}(\cdot)$  denote, respectively, the expectation and variance of a RV. Additionally, $f_X(x)$ and $F_X(x)$ denote the probability density function (PDF) and cumulative distribution function (CDF) of the RV, $X$, respectively. {$\Pr(\cdot)$ denotes the probability of an event.}

\section{System Model}
\begin{figure}[htpb]
\centering
\includegraphics[width=0.9\linewidth]{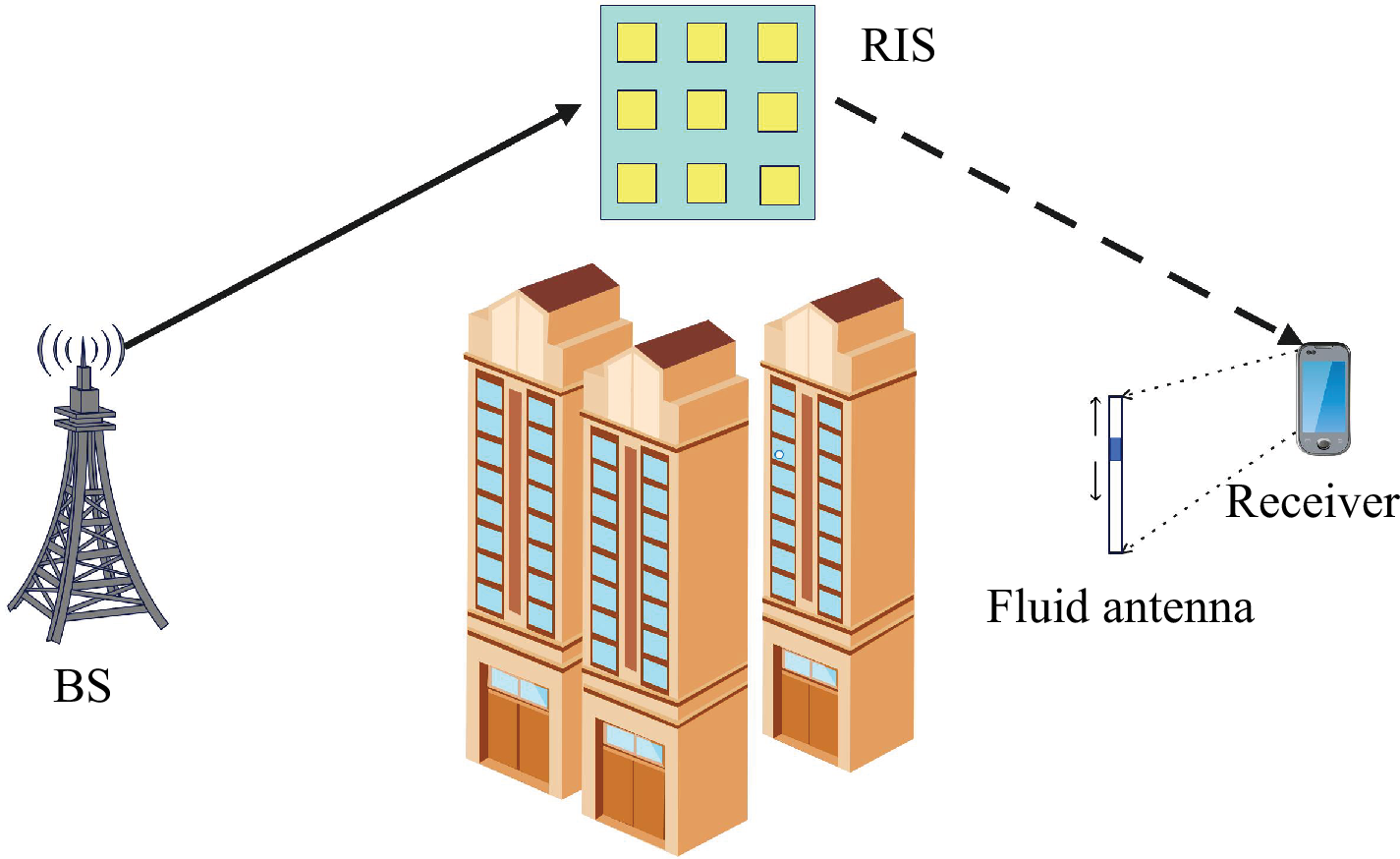}\\
\caption{A RIS-assisted communication system with a FAS-enabled user.}\label{fig1}
\end{figure}

As depicted in Fig.~\ref{fig1}, a RIS-assisted communication system is considered. The communication system consists of a base station (BS) equipped with a single fixed-position antenna, a RIS with $M$ reflecting elements, and a receiver equipped a fluid antenna capable of switching among $N$ ports within a linear space of $W\lambda$, where $W$ is the normalized size, and $\lambda$ is the carrier wavelength. {Here the direct link between the BS and the receiver is assumed broken, due to blockages such as massive buildings or trees, and therefore the RIS is employed to help the transmission of signal from the BS to the receiver.} It is also assumed that the time delay for port switching is negligibly small.

{Some other kinds of smart antenna techniques in table I, such as extremely large-scale (XL) multiple-input multiple-output (MIMO) and movable antenna (MA), can also be applied to improve the system performance \cite{SA24,Yang24}. However, the FAS is more suitable for space-limited  devices, such as receivers in vehicular application scenarios. Hence we consider the FAS for receiver in this paper.}
\begin{table}[h!t]\label{ta1}
\center
\caption{{Comparison to State-of-the-Art antenna techniques}}
\begin{tabular}{m{50pt}<{\raggedright}m{40pt}<{\raggedright}m{50pt}<{\raggedright}m{70pt}<{\raggedright}}
ine\\[-2.9mm]ine
Nomenclature & Position &Receiver size & Number of radiating elements \\
ine
XL-MIMO & Fixed & Large & Massive\\
MA& Flexible &Medium&Few\\
FAS& Flexible &Tiny &One or few\\
ine\\[-2.9mm]ine
\end{tabular}
\end{table}

\subsection{Communication Model}
{In the communication progress, the BS emits the RF signal $s\sim\mathcal{CN}(0,1)$, and then the RIS reflects the transmitted signal to the receiver. In this work, the Rayleigh fading channel is considered, which is suitable for non-light-of-sight scenarios, such as BS-to-vehicle scenarios and vehicle-to-vehicle scenarios.}
Given the absence of the direct link between the  BS and the receiver, the receiver relies on the signals solely from the reflections of the RIS. Consequentially, at the $k$-th FAS port of the user, the received signal is given by \cite{Yang20}
\begin{align}\label{a1}
y_{k}=\sqrt{P_S}\sum_{m=1}^{M}h_m v_{m,k}e^{-2i\pi\theta_m }s+n_k,
\end{align}
where $h_m\sim\mathcal{CN}(0,\epsilon_1)$ denotes the complex channel coefficient from the BS to the $m$-th RIS element, and $v_{m,k}\sim\mathcal{CN}(0,\epsilon_2)$ denotes the channel from the $m$-th RIS element to the $k$-th FAS port, {$\epsilon_1$ and $\epsilon_2$ are the average channel gain of the BS-to-RIS link and RIS-to-receiver link, respectively.} $P_S$ is the transmit power, $n_k\sim\mathcal{CN}(0,\sigma^2_n)$ is the additional white Gaussian noise, $\theta_m$ denotes the reflecting phase of the $m$-th RIS element, and the RIS is assumed to have full knowledge of the phases of $h_m$ and $v_{m,k}$; thus can adjust $\theta_m$ to maximize the channel gain, i.e., $\theta_m=-\arg (h_m v_{m,k})$. Follow this, the combined channel at the $k$-th FAS port can be expressed as
\begin{align}\label{a2}
\gamma_k=\sum_{m=1}^{M}|h_m| |v_{m,k}|.
\end{align}
It can be readily observed that both $|h_m|$ and $ |v_{m,k}|$ follow Rayleigh distribution, $\mathbf{E}(|h_m| |v_{m,k}|)=\mathbf{E}(|h_m|)\mathbf{E}(|v_{m,k}|)=\pi\sqrt{\epsilon_1\epsilon_2}/4$, and $\mathbf{Var}(|h_m| |v_{m,k}|)=\mathbf{E}(|h_m|^2 |v_{m,k}|^2)-(\mathbf{E}(|h_m| |v_{m,k}|))^2=\epsilon_1\epsilon_2(1-\pi^2/16)$.

At the FAS, the port with the maximum channel gain will be selected for receiving the signal. Consequentially, the resulting channel at the optimal port is written as \footnote{In this paper, only one port is activated with for signal processing. When multiple ports are activated, the maximum ratio combining technique can be applied, while this incurs significant implementation complexity \cite{XLai23}.}
\begin{align}\label{a3}
\gamma^*=\max_{k\in\{1,2,\dots,N\}}\gamma_{k}.
\end{align}
{
\begin{remark}
The signal processing method in FAS resembles the selection combining technique in multiple-antenna systems. However, our distinctive contribution is to analyze the system performance including the correlation among the ports.
\end{remark}
}

\subsection{FAS Channel Correlation Model}
For the correlation between two different ports of the FAS, the 3D Clarke's model is adopted\footnote{In this work, the 3D Clarke's model is considered, which is more accurate in practice. Other kinds of model, such as 2D Jake's model can be considered and the derivations in this work are still valid with minor modifications. }. Specifically, the correlation coefficient between ports $k$ and $l$ is modeled by
\begin{align}\label{a4}
g_{k,l}=\mathrm{sinc}\left(\frac{2\pi (k-l) W}{N-1}\right),
\end{align}
where $\mathrm{sinc}(x)=\frac{\sin(x)}{x}$. Note that $\mathrm{sinc}(x)$ is an odd function, which indicates that the correlation coefficient matrix of $\mathbf{v}_m=[v_{m,1},\dots,v_{m,N}]$ is a Toeplitz matrix as
\begin{align}\label{a5}
\mathbf{\Sigma}\in \mathbb{R}^{N\times N} & =\mathbf{topelitz}(g_{1,1},g_{1,2},\dots, g_{1,N})\notag\\
&=\left(
         \begin{array}{cccc}
         g_{1,1} &g_{1,2} & \cdots & g_{1,N} \\
        g_{1,2} & g_{1,1} & \cdots & g_{1,N-1} \\
        \vdots &  & \ddots & \vdots \\
       g_{1,N}& g_{1,N-1} & \cdots &g_{1,1} \\
         \end{array}
       \right).
\end{align}

\section{Outage Performance Analysis}
In this section, the outage performance of the system is analyzed. First, let us denote $R$ as the target transmission rate, and the outage probability is accordingly given as
\begin{align}\label{b1}
P_{\rm{out}}&=F_{\gamma^*}(\Lambda_{\rm{th}})=\Pr\left(\log_2\left(1+\frac{P_S(\gamma^*)^2}{\sigma_n^2}\right)\leq R\right),
\end{align}
where $\Lambda_{\rm{th}}=((2^R-1)\sigma_n^2/P_S)^{1/2}$. To proceed with the outage probability analysis, it is necessary to derive the CDF of $\gamma^*$, i.e., $F_{\gamma^*}(y)$. Given the fact that the exact expression for $F_{\gamma^*}(y)$ is intractable,  the following subsections will develop several approximations for $F_{\gamma^*}(y)$. {Besides, it should be noted that the results presented are developed based on the CLT, and a multivariate Gaussian distribution is assumed. Consequently, the mean and variance of $\gamma_k$, as well as the correlation coefficient between ports $k$ and $l$, must be computed to facilitate further derivations.}

\subsection{CLT Approximation}
In practice, the number of RIS elements, $M$, is supposed to be very large, and thus the CLT can be applied. 
Consequentially, $\mathbf{\Gamma}=[\gamma_1, \dots, \gamma_N]$ can be approximated by $\mathbf{\bar{\Gamma}}=[\bar{\gamma}_1, \dots, \bar{\gamma}_N]$, which follows a multivariate Gaussian distribution with identical correlation coefficient matrix and mean vector  to those of $\mathbf{\Gamma}$. Also, the PDF and CDF of $\mathbf{\Gamma}$ are similar to those of $\mathbf{\bar{\Gamma}}$. 
Furthermore, the mean and variance of $\bar{\gamma}_k$ are identical to those of $\gamma_k$, which can be computed as
\begin{align}\label{c1}
E_{\gamma}=&M\mathbf{E}(|h_m| |v_{m,k}|)=\frac{M\pi\sqrt{\epsilon_1\epsilon_2}}{4},\\ \label{c2}
V_{\gamma}=&M\mathbf{Var}(|h_m| |v_{m,k}|)=M\epsilon_1\epsilon_2\big(1-\frac{\pi^2}{16}\big).
\end{align}
Moreover, as $\gamma_k$ for $k=1,\dots,N$ share the same $|h_m|$ for $m=1,\dots,M$, and $|v_{m,k}|$ are correlated for different $k$, it can be easily demonstrated that $\bar{\gamma}_k$ for $k=1,\dots,N$ are correlated. {Hence, given the calculation of Pearson correlation coefficient and (\ref{a2}), the correlation coefficient between $\gamma_k$ and $\gamma_l$, which indicates the degree of correlation between RVs $\gamma_k$ and $\gamma_l$ and  falls in the range of $[-1,1]$, can be formulated as}
\begin{equation}\label{c3}
\eta(g_{k,l})=\frac{\mathbf{E}(\gamma_k\gamma_l)-E_{\gamma}^2}{V_{\gamma}}=\frac{M\epsilon_1\mathbf{E}(|v_{m,k}||v_{m,l}|)-E_{\gamma}^2/M}{V_{\gamma}},
\end{equation}
where
\begin{align}\label{c4}
\mathbf{E}(|v_{m,k}||v_{m,l}|)=\int_{0}^{\infty}\int_{0}^{\infty}xyf_{|v_{m,k}|,|v_{m,l}|}(x,y)dxdy,
\end{align}
and 
\begin{multline}\label{c5}
f_{|v_{m,k}|,|v_{m,l}|}(x,y)=\frac{4x^2y^2}{\epsilon_2^2(1-g_{k,l})}e^{-\frac{x^2}{\epsilon_2}}e^{-\frac{y^2+g_{k,l}x^2}{\epsilon_2^2(1-g_{k,l})}}\\
\times I_0\left(\frac{2g_{k,l}xy}{\epsilon_2(1-g_{k,l})}\right),
\end{multline}
in which $f_{|v_{m,k}|,|v_{m,l}|}(x,y)$ is the joint PDF of RVs $|v_{m,k}|$ and $|v_{m,l}|$, $I_0(\cdot)$ denotes the zeroth order of Bessel function of the first kind. Note that $\mathbf{E}(|v_{m,k}||v_{m,l}|)$ can be computed via numerical integration.

Based on the correlation coefficient in \eqref{c3}, we can formulate the correlation coefficient matrix of $\mathbf{\bar{\Gamma}}$ as
\begin{align}\label{c6}
\mathbf{\Omega}\in\mathbb{R}^{N\times N}=\mathbf{toeplitz}\left(\eta(g_{1,1}),\dots,\eta(g_{1,N})\right),
\end{align}
and the joint  PDF of $\mathbf{\bar{\Gamma}}$ is written as
\begin{align}\label{c7}
f_{\gamma_1,\dots,\gamma_N}(x_1,\dots,x_N)=\frac{e^{-\frac{1}{2}(\mathbf{X}-\mathbf{E}_{\gamma})^{T}(V_{\gamma}\cdot\mathbf{\Omega})^{-1} (\mathbf{X}-\mathbf{E}_{\gamma})}}{(2\pi V_{\gamma})^{N/2}(\mathrm{det}\mathbf{\Omega} )^{1/2}},
\end{align}
where
\begin{align}\label{c8}
\mathbf{X}=[x_1,\dots,x_N]^{T},~\mbox{and}~\mathbf{E}_{\gamma}=[E_{\gamma}, \dots, E_{\gamma}]^T.
\end{align}
Utilizing the PDF of $\mathbf{\bar{\Gamma}}$ in \eqref{c7}, the CDF of RV, $\bar{\gamma}^*$, can be expressed via the joint CDF of $N$-dimensional Gaussian distribution, i.e., $\Phi(y,V_{\gamma}\cdot\mathbf{\Omega},E_{\gamma})$, as
\begin{align}\label{c9}
&F_{\bar{\gamma}^*}(y)\notag\\
&=\Phi(y,V_{\gamma}\cdot\mathbf{\Omega},E_{\gamma})\notag\\
&= \underbrace{\int_{-\infty}^{y}\cdots\int_{-\infty}^{y}}_{N}f_{\gamma_1,\dots,\gamma_N}(x_1,\dots,x_N)dx_1\cdots dx_N,
\end{align}
where the calculation of $\Phi(y,V_{\gamma}\cdot\mathbf{\Omega},E_{\gamma})$ can be implemented via mathematical tools, such as Matlab and Maple. However, for large values of $N$, say $N\geq 25$, the complexity of the calculation of $F_{\bar{\gamma}^*}(y)$ is prohibitive as an $N$-dimensional integral is required. To overcome the difficulty, we will provide approximate expressions of $F_{\bar{\gamma}^*}(y)$ in the next subsections.

\subsection{Block-Correlation Channel Approximation}
In this part, we resort to a block-correlation approximation model in \cite{BC24}. {It is important to note that as the dimension $N$ increases, the correlation coefficient matrix becomes predominantly influenced by a small fraction of eigenvalues. To efficiently capture the significant eigenvalues, we employ a block matrix, which simplifies the representation of the original matrix.} {Consequently, we derive an approximate distribution for the FAS model, characterized by a correlation coefficient matrix configured in block form. The number of blocks and the sizes of each block are determined based on the principal eigenvalues of ${\bf\Sigma}$.}
Specifically, the random vector $\mathbf{v}_m$ can be approximated by $\bar{\mathbf{v}}_m$, and the correlation coefficient matrix of $\bar{\mathbf{v}}_m$ can be presented in the following form:
\begin{align}\label{d1}
\mathbf{\hat{\Sigma}}\in\mathbb{R}^{N\times N}=\left(
         \begin{array}{cccc}
         \mathbf{A}_1 & \mathbf{0} & \cdots & \mathbf{0} \\
        \mathbf{0} & \mathbf{A}_2 & \cdots & \mathbf{0} \\
        \vdots &  & \ddots & \vdots \\
        \mathbf{0} & \mathbf{0} & \cdots & \mathbf{A}_D \\
         \end{array}
       \right),
\end{align}
   where
\begin{align}\label{d2}
\mathbf{A}_d\in\mathbb{R}^{L_d\times L_d}=\left(
         \begin{array}{cccc}
         1 & \mu & \cdots & \mu \\
        \mu & 1 & \cdots & \mu \\
        \vdots &  & \ddots & \vdots \\
        \mu & \mu& \cdots & 1 \\
         \end{array}
       \right),
\end{align}
$\mu$ denotes a constant close to $1$, and $\sum\nolimits_{d=1}^{D}L_d=N$ with $D=|S\{\lambda_{th}\}|$ being determined by the number of principal eigenvalues of ${\bf\Sigma}$. That is, there exists $S\{\lambda_{\rm{th}}\}=\{\lambda_n|\lambda_n\geq\lambda_{\rm{th}}, n=1,\dots, N\}$, and $D$ equals to the cardinal number of $S\{\lambda_{th}\}$, where $\lambda_{\rm{th}}$ is a small number to ensure that enough eigenvalues are included in $S\{\lambda_{\rm{th}}\}$, which can be dynamically adjusted. {After identifying the principal eigenvalues of ${\bf\Sigma}$, which dictate the number of blocks, we proceed to ascertain the size of each block.} {Specifically, $L_d$ represents the size of the $d$-th block, which is determined by}
\begin{align}\label{d3}
\underset{L_1,\dots,L_d,\dots,L_D}{\arg\min}~~\mathrm{dist}(\hat{\bf\Sigma},{\bf\Sigma}),
\end{align}
where $\mathrm{dist}(\cdot)$ is a distance metric between two matrices, which is determined by the difference of their eigenvalues and the detailed procedure can be found in \cite{BC24}.
{ {The procedure outlined in reference \cite{BC24} involves determining the integer $L_d$ for each block sequentially, based on \eqref{d3}. This process continues until  $\sum_{d=1}^D L_d=N$, ensuring that the sum of the eigenvalues of the approximated matrix $\hat{\bf\Sigma}$ matches that of the original matrix ${\bf\Sigma}$.} }

Based on the block-correlation model, we can approximate $\mathbf{\Gamma}$ with $\hat{\mathbf{\Gamma}}=[\hat{\gamma}_{1}, \dots, \hat{\gamma}_N]$. Applying the Pearson correlation coefficient calculation,  the correlation coefficient matrix of $\hat{\mathbf{\Gamma}}$ is given as
\begin{align}\label{d4}
\mathbf{\hat{\Omega}}\in\mathbb{R}^{N\times N}=\left(
         \begin{array}{cccc}
         \mathbf{B}_1 & \mathbf{C} & \cdots & \mathbf{C} \\
        \mathbf{C} & \mathbf{B}_2 & \cdots & \mathbf{C} \\
        \vdots &  & \ddots & \vdots \\
        \mathbf{C} & \mathbf{C} & \cdots & \mathbf{B}_D \\
         \end{array}
       \right),
\end{align}
where all the elements in the submatrix within $\mathbf{C}$ are given as $\ra=\eta(0)$, and
\begin{align}\label{d5}
\mathbf{B}_d\in\mathbb{R}^{L_d\times L_d}=\left(
         \begin{array}{cccc}
         1 & \rb & \cdots & \rb \\
        \rb & 1 & \cdots & \rb \\
        \vdots &  & \ddots & \vdots \\
       \rb & \rb& \cdots & 1 \\
         \end{array}
       \right),
\end{align}
where $\rb=\eta(\mu)$.

Based on the structure of the correlation coefficient matrix $\hat{\mathbf{\Omega}}$, which indicates that the elements in $\hat{\mathbf{\Gamma}}$  can be divided into $D$ parts, and each part contains $L_d$ identically distributed RVs, for $d\sd$, we can then rearrange the subscript and rewrite $\hat{\mathbf{\Gamma}}$ into $[\hat{\gamma}_{1,1},\dots,\hat{\gamma}_{L_1,1},\dots,\hat{\gamma}_{k,d},\dots,\hat{\gamma}_{L_D,D}]$. 
{Thanks to the block-correlation model, where the correlation coefficient within each block is a constant, the analysis becomes tractable for the FAS-RIS communication and other complicated FAS-enabled communications.}
Here, $\hat{\gamma}_{k,d}$ can be reformulated as
\begin{align}\label{d6}
\hat{\gamma}_{k,d}=&\sqrt{1-\rb}z_{2,k,d}+\sqrt{\rb-\ra}z_{1,d}+\sqrt{\ra}z_0+E_{\gamma}\notag\\
=&a_{k,d}+r_{d}+w,
\end{align}
for $k=1,\dots,L_d$, and $d=1,\dots,D$, and $z_{2,k,d}$, $z_{1,d}$ and $z_0$ are independently and identically distributed (i.i.d.) Gaussian RVs with mean $0$ and variance $V_{\gamma}$. Also, $w\sim\mathcal{N}(E_{\gamma},V_{\gamma}\ra)$,  $r_d\sim\mathcal{N}(0,V_{\gamma}(\rb-\ra))$ and
 $a_{k,d}\sim\mathcal{N}(0,V_{\gamma}(1-\rb))$ for $k=1,\dots,L_d$ and $d=1,\dots,D$, are independently distributed.

The maximum of the channel parameter $\hat{\gamma}_{k,d}$ in each block is $\hat{\gamma}_d^*=\max_{k\in\{1,\dots,L_d\}} \hat{\gamma}_{k}^*$ for $d=1,\dots,D$, and the CDF of RV $\hat{\gamma}^*=\max_{d\in\{1,\dots,D\}} \hat{\gamma}_{d}^*$ can be computed as
\begin{align}\label{d7}
F_{\hat{\gamma}^*}(y)=\mathbf{E}_w\left[\prod_{d=1}^{D}F_{\hat{\gamma}_d^*|w}(y)\right].
\end{align}

Applying the PDFs of RVs $a_{k,d}$, $r_d$ and $w$, that is
\begin{align}\label{d8}
f_{a_{k,d}}(x)&=\frac{1}{\sqrt{2\pi V_{\gamma}(1- \rb )}}e^{-\frac{x^2}{2V_{\gamma}(1- \rb )}}, \\
f_{r_d}(x_d)&=\frac{1}{\sqrt{2\pi V_{\gamma}( \rb - \ra )}}e^{-\frac{x_d^2}{2V_{\gamma}( \rb - \ra )}}, \\
f_{w}(x_0)&=\frac{1}{\sqrt{2\pi V_{\gamma} \ra }}e^{-\frac{(x_0-E_{\gamma})^2}{2 V_{\gamma} \ra }},
\end{align}
we can obtain the conditional PDF of $\hat{\gamma}_{k,d}$ for $k=1,\dots,L_d$ and $d=1,\dots,D$ with $r_d=x_d$ and $w=x_0$ as
\begin{align}\label{d9}
f_{\hat{\gamma}_{k,d}|r_d=x_d,w=x_0}(x)=\frac{e^{-\frac{(x-x_d-x_0))^2}{2V_{\gamma}(1- \rb )} }}{\sqrt{2\pi V_{\gamma}(1- \rb )}},
\end{align}
 and the conditional CDF as
\begin{align}\label{d10}
&F_{\hat{\gamma}_{k,d}|r_d=x_d,w=x_0}(y)\notag\\
&=\int_{-\infty}^{y}f_{\gamma_{k,d}|r_d=x_d,w=x_0}(x)dx\notag\\
&=\frac{1}{2}\left[1+\mathrm{erf}\left(\frac{y-x_d-x_0)}{\sqrt{2\pi V_{\gamma}(1- \rb )}}\right)\right],
\end{align}
where ${\rm erf}(x)$ is the error function. Taking the expectation of $F_{\gamma_{k,d}|r_d=x_d,w=x_0}(y)$ over RV $r_d$, we obtain
\begin{align}\label{b21}
F_{\hat{\gamma}_d^*|w=x_0}(y)=\int_{-\infty}^{\infty}\big(F_{\gamma_{k,d}|r_d=x_d,w=x_0}(y)\big)^{L_d}f_{r_d}(x_d)dx_d.
\end{align}
Finally, we can achieve
\begin{align}\label{b22}
&F_{\hat{\gamma}^*}(y)\notag\\
&=\int_{-\infty}^{\infty}\prod_{d=1}^{D}F_{\hat{\gamma}_d^*|w=x_0}(y)f_{w}(x_0)dx_0\notag\\
&=\int_{-\infty}^{\infty}\prod_{d=1}^{D}\int_{-\infty}^{\infty}\frac{1}{2^{L_d}}\left[
1+\mathrm{erf} \left(\frac{y-x_d-x_0}{\sqrt{2\pi V_{\gamma}(1- \rb )}} \right)\right]^{L_d}\notag\\
&\quad\times f_{r_d}(x_d)dx_d f_{w}(x_0)dx_0\notag\\
&\approx \frac{H\pi}{U^2V_{\gamma}}\sum_{l=1}^{U}\prod_{d=1}^{D}\sum_{t_{d}=1}^{U}\frac{\left[1+\mathrm{erf} \left(\frac{y-H p_{t,d}-Hq_l}{\sqrt{2\pi V_{\gamma}(1- \rb )}} \right)\right]^{L_d}}{2^{L_d+1}}\notag\\
&\quad\times\sqrt{\frac{(1-p_{t,d}^2)(1-q_{l}^2)}{ \ra( \rb - \ra )}}
e^{-\frac{(H p_{t,d})^2}{2V_{\gamma}( \rb - \ra )}-\frac{(Hq_l-E_{\gamma})^2}{2 V_{\gamma} \ra }},
\end{align}
where the Gauss-Chebyshev integral is applied in the approximation of the final step \cite{NumericalAnalysis}, $H$ is a large number, $U$ is a number to tradeoff accuracy and complexity\footnote{For $D$-fold Gauss-Chebyshev integral, the computational complexity is $\mathcal{O}(U^N)$, and the degree of algebraic precision is $2U-1$.}, and
\begin{align}\label{b23}
p_{t,d}=\cos\left(\frac{(2t_d-1)\pi}{2U}\right),~\mbox{and}~q_{l}=\cos\left(\frac{(2l-1)\pi}{2U}\right).
\end{align}

\subsection{CLT and I.I.D.~Channel Approximation}
To gain further insight, we consider the setting that $\mu$ is close to 1 with $\mu\rightarrow 1$, and hence $\eta(\mu)\rightarrow 1$, and the block-correlation channels can be viewed as $D$ i.i.d.~channels. Thus, in the approximated channel model, we have
\begin{align}\label{e1}
\tilde{\gamma}_k=\sqrt{1- \ra }b_k+\sqrt{ \ra }b_0+E_{\gamma},
\end{align}
where $b_k\sim\mathcal{N}(0,V_{\gamma})$ for $k=0,1,\dots,D$. Also, the PDF of $b_k$ is given by
\begin{align}\label{e2}
f_{b_k}(x)=\frac{1}{\sqrt{2\pi V_{\gamma}}}e^{-\frac{x^2}{2V_{\gamma}}}, k=0,1,\dots,D.
\end{align}
Accordingly, we can write the conditional PDF of $\tilde{\gamma}_k$ with $b_0=x_0$ as
\begin{align}\label{e3}
f_{\tilde{\gamma}_k|b_0=x_0}(x)=\frac{e^{-\frac{(x-(E_{\gamma}+ \ra x_0))^2}{2V_{\gamma}(1- \ra )}}}{\sqrt{2\pi V_{\gamma}(1- \ra )}},
\end{align}
and correspondingly, the conditional CDF becomes
\begin{align}\label{e4}
F_{\tilde{\gamma}_k|b_0=x_0}(y)&=\int_{-\infty}^{y}
\frac{e^{-\frac{(x-(E_{\gamma}+ \ra x_0))^2}{2V_{\gamma}(1- \ra )}}}{\sqrt{2\pi V_{\gamma}(1- \ra )}}dx\notag\\
&=\frac{1}{2}\left[1+{\rm erf}\left(\frac{y-(E_{\gamma}+ \ra x_0)}{\sqrt{2V_{\gamma}(1- \ra )}} \right)\right].
\end{align}
Moreover, taking the expectation of $F_{\tilde{\gamma}_k|b_0=x_0}(y)$ over RV $b_0$, we can write the CDF of $\tilde{\gamma}^*$ as
\begin{align}\label{e5}
F_{\tilde{\gamma}^*}(y)=&\int_{0}^{\infty}\big[F_{\tilde{\gamma}_k|b_0=x_0}(y)\big]^{D} f_{b_0}(x_0)dx_0\notag\\
\approx&\frac{H\pi}{U}\sum_{l=1}^U \frac{1}{2^D}\left[1+{\rm erf}\left(\frac{y-E_{\gamma}- \ra Hq_l}{\sqrt{2V_{\gamma}(1- \ra )}} \right)\right]^D \notag\\
&\times \sqrt{\frac{1-q_l^2}{2\pi V_{\gamma}}}e^{-\frac{(Hq_l)^2}{2V_{\gamma}}}.
\end{align}

\begin{remark}
As can be observed, the approximate CDF of $\gamma^*$ in \eqref{c9} requires $N$-fold integration, while the approximate CDF in \eqref{b22} and \eqref{e5} require $(D+1)$-fold integration and one-fold integration, respectively. This indicates that the block-correlation based approximation can be effective in evaluating the performance of the FAS-RIS system, for large $N$.
\end{remark}

\begin{remark}
From \eqref{b22} and \eqref{e5}, we can see that the outage probability decreases with a larger $D$; yet the value of $D$ may not decrease linearly with the number of FAS ports $N$ and $D\ll N$ when the value of $N$ is huge \cite{BC24}. Therefore, it is not surprising that the enhancement brought by increasing $N$ become less evident when $N$ is large.
\end{remark}

\section{Numerical Results}
In this section, we provide numerical results for evaluating the outage performance of the FAS-RIS system. The distance between the BS and RIS and that between the RIS and the mobile receiver are both set to $200$ meters. For path-loss factor of $2$, we have $\epsilon_1=\epsilon_2=200^{-2}$. Also, $W=1$ except Fig.~\ref{fig3}. {The accuracy-complexity tradeoff number $U$ is set to 100.}

\begin{figure}[htbp]
\centering
\includegraphics[width=0.8\linewidth]{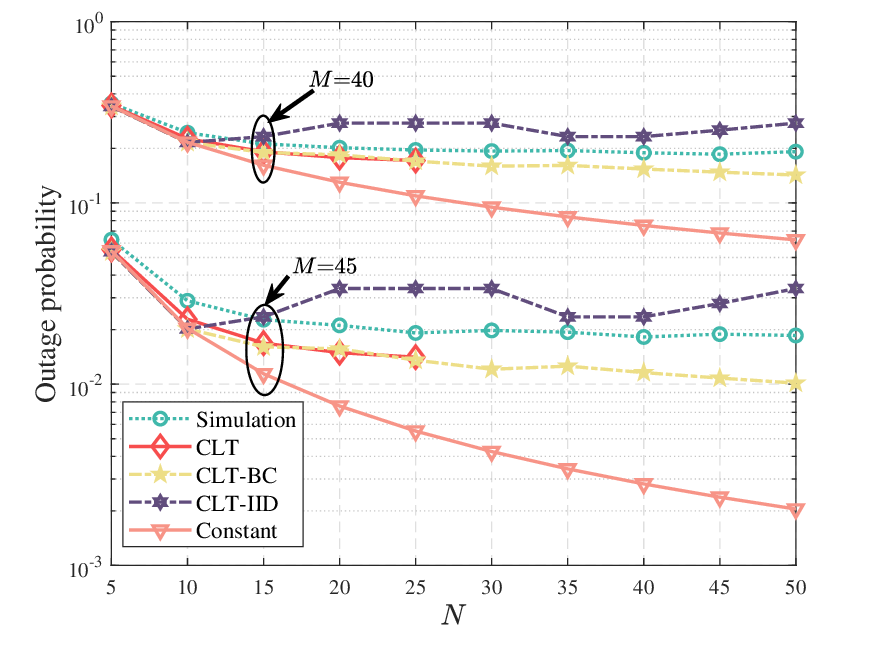}\\
\caption{Outage probability versus the number of ports $N$. }\label{fig2}
\end{figure}

\begin{figure}[htbp]
\centering
\includegraphics[width=0.72\linewidth]{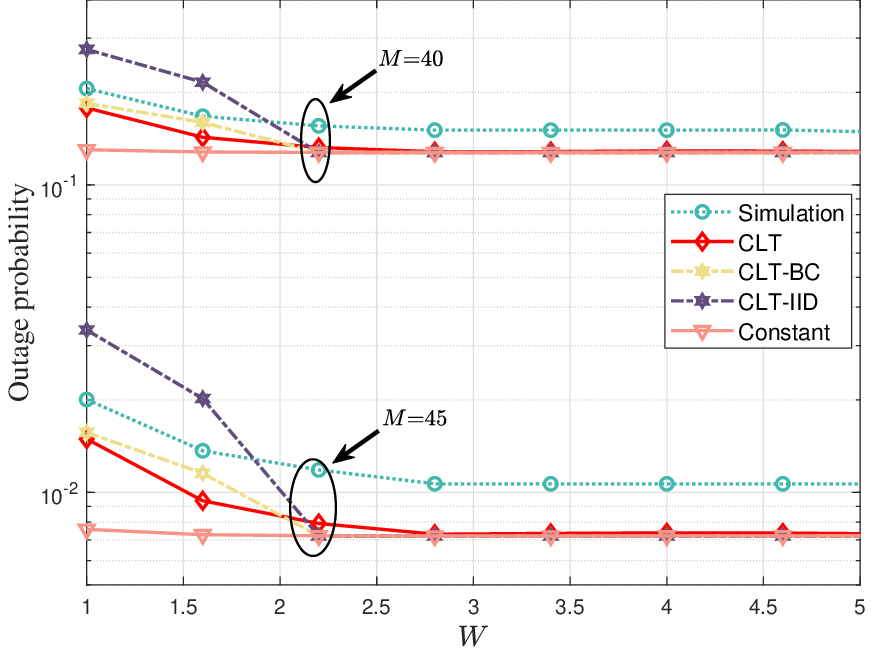}\\
\caption{Outage probability versus normalized size $W$.}\label{fig3}
\end{figure}

In Fig.~\ref{fig2}, the variations of outage probability with different values of the number of ports $N$ are observed, where $M=40,45$ and $N$ varies from $5$ to $50$. The noise power is $\sigma_n^2=10^{-8}~{\rm W}$, and the transmit power is $P_S=0.1~{\rm W}$. Hence the average  received  SNR are $\mathbf{E}(P_S|\gamma_k|^2/\sigma^2)=10$ dB and $11.0231$ dB for $M=40$ and $M=45$ respectively. The target data rate $R=3~{\rm bit/s/Hz}$. In the numerical results, we give the results based on \eqref{c9}, \eqref{b22}, and \eqref{e5}, which are denoted as ``CLT", ``CLT-BC", ``CLT-IID" respectively. For the block correlation model, $\mu$ is set to $0.9$. For comparison, the results based on the constant correlation model in \cite{FAS21} are also given and denoted as ``Constant". Note that \cite{FAS21} is not accurate. 
{Also, when $W$ is large or $N$ is small, the correlation among the ports becomes negligible, which matches the assumption of the constant correlation model. Hence the results of constant correlation model  and the simulation results are close.}

As observed, the block-correlation model based results align well with the CLT based results, which suggests that the block-correlation model can help avoid the heavy computation of the CDF of multi-dimensional Gaussian distribution to analyze the performance of FAS-RIS systems. Also, we see that the outage probability decreases with a larger value of $N$, especially when the value of $N$ is small. When the value of $N$ is large, such as $N\geq 30$, the outage probability remains unchanged with different $N$. This phenomenon can also be observed from the results of the block-correlation model. However, the results of constant correlation model are much more lower than other results when the value of the number of ports $N$ is large, which suggests that the results of constant correlation model might overestimate the system performance.

%

In Fig.~\ref{fig3}, the results are provided for the outage probability with different values of $W$, where $M=40,45$, $N=20$ and $W$ varies from $1$ to $5$. As observed, with larger values of $W$, the outage probability decreases, which is more obvious when $W\leq 3$. This is because when the value of $W$ increases, the statistical dependency among the channel parameters of each port decreases, and more effective diversity can be achieved for the FAS-enabled receiver.

\begin{figure}[htbp]
\centering
\includegraphics[width=0.8\linewidth]{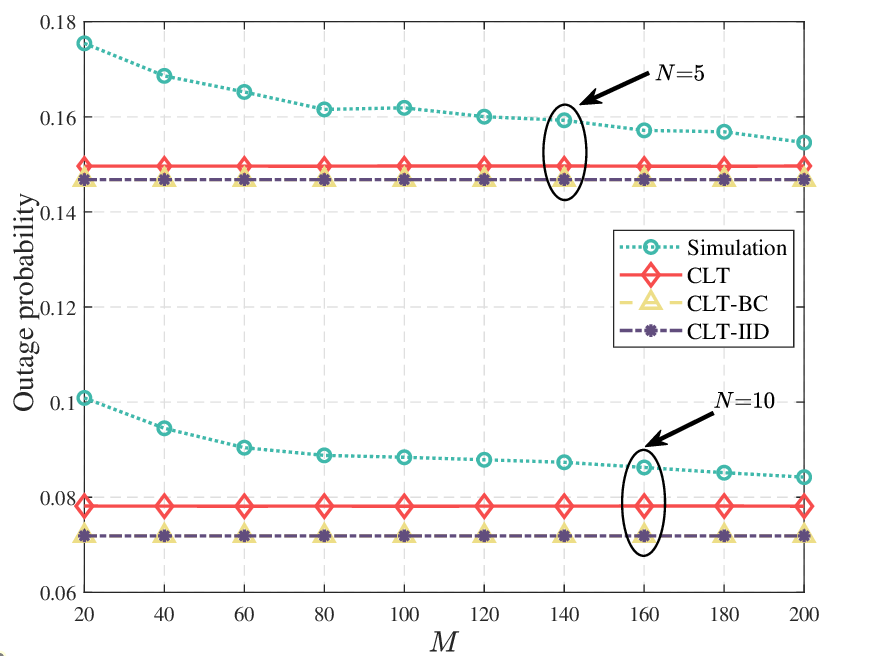}\\
\caption{Outage probability versus the number of RIS elements $M$.}\label{fig4}
\end{figure}

In Fig.~\ref{fig4}, we plot the outage probability of FAS-RIS with different values of $M$, where $M$ varies from $20$ to $200$, and $N=5,10$. To better demonstrate the impact of CLT on the analytical results, we scale the outage threshold $\Lambda_{\rm{th}}$ to $E_{\gamma}$. As observed, the accuracy of the CLT based approximation results and the block-correlation based approximation results become more satisfying with large values of $M$.

\section{Conclusions}
In this correspondence, we analyzed the outage probability for FAS-RIS systems. Applying the CLT and block-correlation channel model, we derived the expressions of outage probability. Numerical results showed the effectiveness of the proposed analysis, as the derived analytical results become close to the simulation results when the number of RIS elements is large.

\end{document}